 \documentclass[manuscript]{aastex63}

\usepackage[utf8]{inputenc}

\usepackage{amsmath,amssymb,amstext}
\usepackage{booktabs}
\usepackage{multirow}

\usepackage{epstopdf}

\usepackage{amsmath,amssymb,amstext}

\usepackage{natbib}
\shorttitle{Machine learning as a flaring storm warning machine}
\shortauthors{Benvenuto et al.}

\begin{document}


\title{Machine learning as a flaring storm warning machine: \\ Was a warning machine for the September 2017 solar flaring storm possible?}


\author{F. Benvenuto}
\affil{Dipartimento di Matematica, Universit\`a di Genova, Genova, Italy}
\email{benvenuto@dima.unige.it}

\author{C. Campi}
\affil{Dipartimento di Matematica, Universit\`a di Padova, Padova, Italy}
\email{campi@dima.unige.it}

\author{A. M. Massone}
\affil{Dipartimento di Matematica, Universit\`a di Genova and CNR - SPIN, Genova, Italy}
\email{massone@dima.unige.it}

\author{M. Piana}
\affil{Dipartimento di Matematica, Universit\`a di Genova and CNR - SPIN, Genova, Italy}
\email{piana@dima.unige.it}

%
%
%
%

%

\begin{abstract}
Machine learning is nowadays the methodology of choice for flare forecasting and supervised techniques, in both their traditional and deep versions, are becoming the most frequently used ones for prediction in this area of space weather. Yet, machine learning has not been able so far to realize an operating warning system for flaring storms and the scientific literature of the last decade suggests that its performances in the prediction of intense solar flares are not optimal. 

The main difficulties related to forecasting solar flaring storms are probably two. First, most methods are conceived to provide probabilistic predictions and not to send binary yes/no indications on the consecutive occurrence of flares along an extended time range. Second, flaring storms are typically characterized by the explosion of high energy events, which are seldom recorded in the databases of space missions; as a consequence, supervised methods are trained on very imbalanced historical sets, which makes them particularly ineffective for the forecasting of intense flares. 

Yet, in this study we show that supervised machine learning could be utilized in a way to send timely warnings about the most violent and most unexpected flaring event of the last decade, and even to predict with some accuracy the energy budget daily released by magnetic reconnection during the whole time course of the storm. Further, we show that the combination of sparsity-enhancing machine learning and feature ranking could allow the identification of the prominent role that energy played as an Active Region property in the forecasting process.
\end{abstract}





\section{Introduction}

September 2017 was the most flare-productive period of solar cycle 24: between September 6 and September 10, twenty-seven M flares and four X flares were emitted by the Sun, which correspondingly released several powerful coronal mass ejections (CMEs) and bursts of high-energy protons. 

In the one hand, the impact of the September 2017 CMEs on Earth magnetosphere was timely and correctly predicted by the Space Weather Prediction Center of the National Oceanic and Atmospheric Administration (SWPC-NOAA) (see the report currently at the URL https://www.swpc.noaa.gov/sites/default/files/images/u4/01$\%$20Rob$\%$20Redmon.pdf). Between September 4 and September 13 two R2 and three R3 alerts were released concerning radio blackouts and, in the same time range, the service sent one G3 (strong level) and one G4 (severe level) watch for CME-induced geomagnetic storms.  

On the other hand, solar physics \citep{taem88} explains that solar flares are the major trigger of space weather, but if one looks for flare warning systems the situation is not comforting. In fact, we can probably state that there is currently no service for flaring storms prediction analogous to what SWPC-NOAA offers for the prediction of geomagnetic events. Indeed, SolarMonitor at https://www.solarmonitor.org/, which is probably the facility that most comes close to this service, has some limitations concerning the kind of information and computational methods exploited for performing the prediction. Specifically, SolarMonitor relies on rather old historical datasets, which are not necessarily worse than the newest ones but which, for example, do not include the rich information on the magnetic field configuration put at disposal by the magnetograms recorded by the {\em{Helioseismic and Magnetic Imager}} onboard the {\em{Solar Dynamics Observatory}} ({\em{SDO/HMI}}) \citep{scetal12}. Further, SolarMonitor utilizes just McIntosh classification \citep{bletal12,gaetal02,mcetal18} for prediction, which is not necessarily less reliable than other approaches but which, for example, does not account for the potentiality of machine learning. Most importantly, SolarMonitor is not an actual flare warning service, since it computes probabilistic predictions with no imposition of an {\em{a posteriori}} optimal threshold on probabilities. 

Our perspective in the present study was to imagine ourselves back to August 2017 and then to wonder whether, in those days, we could have been able to warn people about the occurrence of solar flares triggering that looming super-storm; of course, this should have been done by using AR observations and data analysis methods that space telescopes and computational science put at our disposal around two years ago. 

As far as AR observations are concerned, since February 2010 {\em{SDO/HMI}} has been providing both line-of-sight and vector magnetograms of the full solar disk at a (vector magnetogram) cadence of $12$ minutes. In particular, for space weather applications, the {\em{Space-Weather HMI Active Region Patches (SHARP)}} process \citep{boetal14} has been developed in order to automatically segment ARs in {\em{HMI}} frames. The Solar Monitor Active Region Tracker (SMART) algorithm \citep{hietal11} has been developed in order to extract active region properties from {\em{SDO/HMI}} line-of-sight magnetograms, while other procedures relying on different methodological approaches have been more recently introduced to determine physical, geometrical, and topological predictors \citep{guetal18,koetal18,paetal17}. Further, a lot of such predictors have been extracted exploiting pattern recognition algorithms developed within the framework of the Horizon 2020 {\em{Flare Likelihood and Region Eruption Forecasting (FLARECAST)}} project (http:flarecast.eu).

As far as computation is concerned, most data analysis approaches in the last decade refer to machine learning as the general framework where to pick up prediction algorithms \citep{ahetal13,bale08,baetal16,beetal18,boco15,coqa09,fletal18,huetal18,lietal07,lietal08,2020ApJ...891...10L,mapi18,mietal17,paetal18,sako17,waetal08,yuetal09}. The assessment of performance for these flare forecasting algorithms is typically made by computing skill scores \citep{bletal12} whose determination relies on confusion matrices. However, if one looks at the values of these scores in the many experiments described so far in the literature, the main surprise is that they are systematically, distinctly (and often significantly) smaller than one, which indicates far from optimal performance \citep{caetal19}. The reasons for these disappointing achievements of machine learning in flare forecasting are probably three. 

First, the largest majority of properties extracted from {\em{HMI}} images include overlapping information and this implies that the datasets of corresponding features are highly redundant for flare prediction \citep{caetal19}. Second, solar flaring storms are seldom and, specifically, flares of class X often characterizing this kind of extreme events are very few in the {\em{HMI}} archive (for example: around 0.3$\%$ of the total amount of X, M, and C flares in the time range between September 2012 and April 2016); this implies that the training sets with which supervised machine learning networks can be optimized are strongly imbalanced against X flares and therefore the corresponding prediction ability is significantly compromised. Third, rather few machine and deep learning methods have been designed so far to exploit the dynamical information contained in AR data and, in any case, their performances typically suffer the imbalanced nature of the training sets \citep{anetal20,huetal18,lietal19,paetal18,paetal20}.

However, the crucial message of the present study is that machine learning based on image features preliminarily extracted from {\em{HMI}} data, may be much more useful to flare forecasting than ever thought so far, even in the case of point-in-time data. Specifically, we show that if point-in-time feature vectors are fed into a regularization network, then this latter can be used as a warning tool able to provide flare/no-flare alerts unrolled along time and that, therefore, if appropriately used, even point-in-time information can be used to realize dynamical predictions. The plan of the paper is as follows. Section 2 describes the forecasting experiment and provide some details about the features utilized in the training/prediction steps and about the data analysis approach. Section 3 shows and discusses the data analysis results. Our conclusions are offered in Section 4.

\section{Methods}
This section describes the prediction experiment performed in the case of the September 2017 flaring storm, with specific focus on the way the training set is constructed, on the features extracted from the {\em{HMI}} images and on the machine learning method utilized in the data processing.
\subsection{Setup of the prediction experiment}
\begin{figure}[h]
\centering
\includegraphics[height=8.5cm]{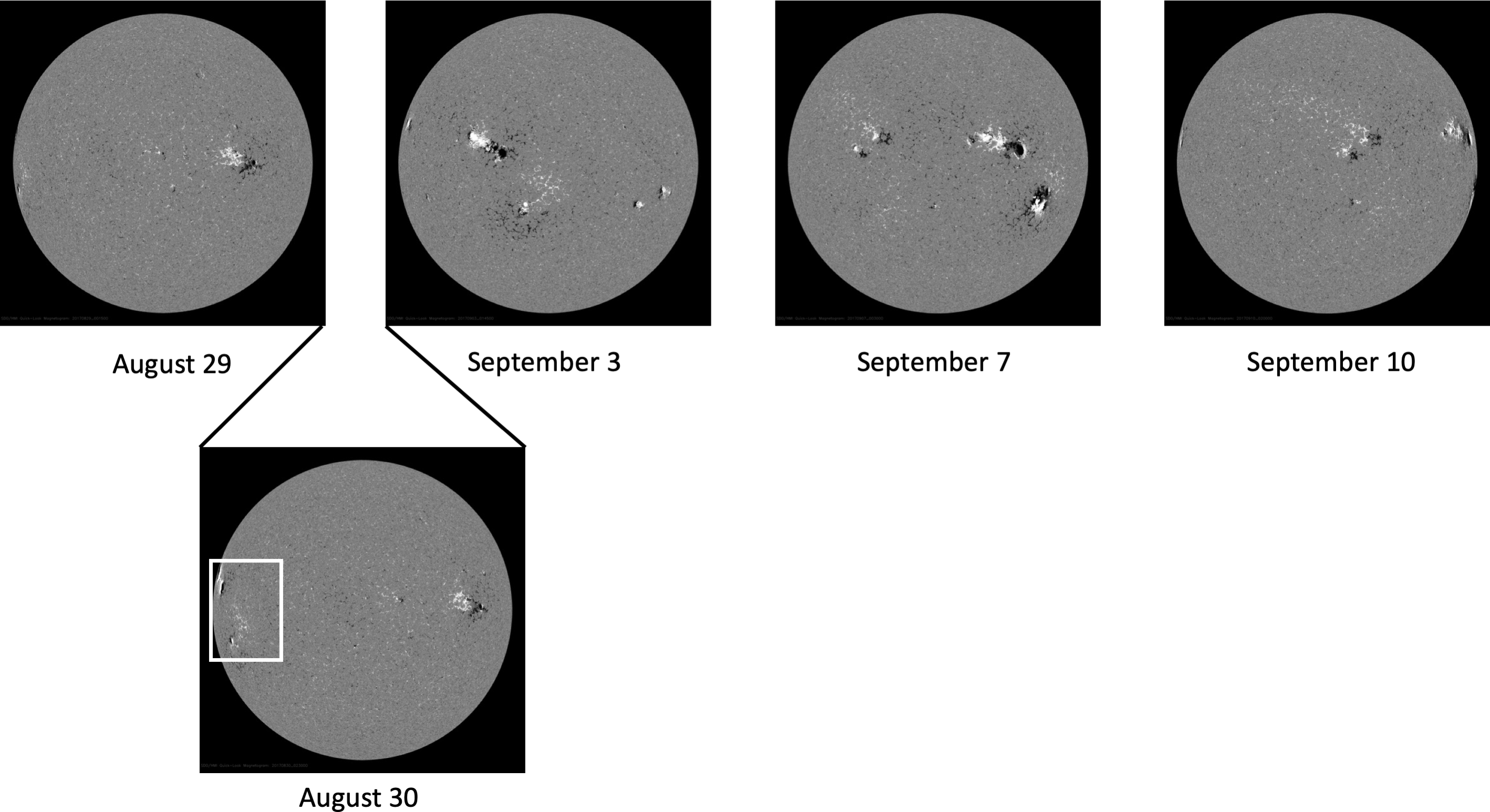}
\caption{A set of {\em{SDO/HMI}} images in the time range between August 29 to September 10 2017. An August 30 frame is highlighted in order to point out the appearance of AR 12673 (in the white box).}
\label{figure:fig-1}
\end{figure}

The forecasting experiment was conceived according to the following steps:
\begin{enumerate}
\item We constructed four sets of {\em{HMI}} images recorded at four issuing times in the range of days between September 14 2012 and April 30 2016. The four issuing times were  00:00 UT, 06:00 UT, 12:00 UT, and 18:00 UT. 
\item We applied pattern recognition \citep{guetal18} in order to extract $19$ features from each one of the ARs in each one of the four sets described in the previous item and we added to them the features representing the longitude and latitude of each AR, a binary feature about flare occurrence in the past $24$ hours and the possible accumulated flare peak magnitude in the past $24$ hours. Then, for each issuing time, we labelled with $1$ each $23$-dimension feature vector for which GOES recorded a flare of GOES class C1 and above (C1+ flares) within the next $24$ hours after the issuing time, and with $0$ each feature vector for which GOES did not record any event in the next $24$ hours. This procedure allowed the construction of four training sets for the C1+ class flares, each one for a specific issuing time. The same procedure was applied in order to construct four training sets for the prediction of flares of class M1 and above (M1+ flares) and of flares of class X1 and above (X1+ flares).
\item A machine learning regularization network for classification was trained by means of the four training sets constructed for the prediction of C1+ flares. Specifically, we utilized the hybrid LASSO method discussed in \citep{beetal18}. This same training step was repeated for the prediction of M1+ and X1+ events.
\end{enumerate}
Figure \ref{figure:fig-1} shows a sample of {\em{HMI}} frames recorded starting from August 29 to September 10 2017. In this sequence, the image corresponding to August 30 has been highlighted in order to point out the appearance of AR 12673, a rather extended AR that was caught by the telescope for some of the following days and which exited from its field-of-view on September 9. Starting from August 30 at 00:00 UT, we began extracting the $23$ features from AR 12673, feeding the corresponding feature vector into the hybrid LASSO algorithm separately trained by using the training sets corresponding to the 00:00 UT issuing time for the prediction of C1+, M1+ and X1+ flares, and annotating whether a flare of corresponding class was predicted in the next $24$ hours. We did the same for the remaining three issuing times and then we continued with the 00:00 UT frame of August 31 and so forth. In this way, machine learning worked as a warning machine characterized by four time windows per day in which to decide whether to send an alert or not, and able to predict three possible lower bounds (C1+, M1+, X1+) for the energetic budget associated to each possible predicted event. We point out that this binary flare/no-flare behavior of the warning machine is made possible by the characteristics of its computational core. Indeed, hybrid LASSO is intrinsically a classification algorithm, which performs an automatic clustering of the LASSO regression outcomes in a way depending on the historical dataset used in the training phase, but not depending on skill scores (i.e., in such a way that no optimization of any specific skill score is needed).

\subsection{Data and data featuers}
Our analysis relied on SHARP data products in the HMI database. These data comprise 2D images of continuum intensity, the full three-component magnetic field vector, and the line-of-sight component of each HARP's photospheric extent. We then made use of property extraction algorithms developed within the FLARECAST Horizon 2020 project to obtain $19$ features; finally, four more properties have been added, corresponding to the longitude and latitude of the AR, a binary label encoding the presence of a flare in the past $24$ hours and the flare index over the past $24$ hours. Therefore the machine learning analysis has been performed against $23$ features overall, i.e.:

\begin{itemize}
\item feature 1: Falconer's gradient-weighted integral length of neutral line;
\item feature 2: heliographic longitude of SHARP centroid;
\item feature 3: heliographic latitude of SHARP centroid; 
\item feature 4: binary flag for occurrence of one flare in previous 24 hr; 
\item feature 5: accumulated GOES flare peak magnitudes in previous 24 hr;  
\item feature 6: total length of all Magnetic Polarity Inversion Lines (MPILs);
\item feature 7: multifractal structure function inertial range index; 
\item feature 8: total unsigned flux around all MPILs; 
\item feature 9: Schrijver's R (log10 form); 
\item feature 10: sum of the horizontal magnetic gradient; 
\item feature 11: Ising energy (calculated pixel-by-pixel);
\item feature 12: maximum length of a single MPIL 
\item feature 13: multifractal generalized correlation dimension spectrum; 
\item feature 14: Ising energy (calculated using Beff flux partitions);
\item feature 15: separation distance between the leading and following polarity subgroups;  
\item feature 16: fractal dimension; 
\item feature 17: total of all separate ratios of MPIL lengths to minimum height of critical decay index;
\item feature 18: ratio of MPIL length to minimum height of critical decay index (for longest MPIL);
\item feature 19: maximum ratio of MPIL length to minimum height of critical decay index;
\item feature 20: ratio of MPIL length to minimum height of critical decay index (for MPIL having lowest minimum height of critical decay index);
\item feature 21: effective connected magnetic field strength;
\item feature 22: Fourier power spectral index;
\item feature 23: continuous wavelet transform power spectral index.
\end{itemize}

\subsection{Prediction and feature ranking methods.} 
The machine learning utilized for data analysis in this paper was a hybrid version of the LASSO regression method \citep{beetal18}. 

LASSO methods \citep{ti96} are intrinsically regression methods and therefore they are not originally conceived for applications that require a binary, yes/no response. However, in \citep{beetal18} a threshold optimization is introduced to realize the classification of the LASSO outcomes by means of fuzzy clustering \citep{beetal84}. The idea of hybrid LASSO is therefore to use LASSO in the first step in order to promote sparsity and to realize feature selection. This step provides an optimal estimate of the model parameters and corresponding predicted output; then, in the second step, Fuzzy C-Means is applied for clustering the predicted output in two classes. 
The main advantage of this approach is in the use of fuzzy clustering to automatically classify the regression output in two classes. Indeed, fuzzy clustering identifies flaring/non flaring events with a thresholding procedure that is data-adaptive and completely operator- and skill-scores-independent.

Once the machine learning method has been applied to the input data, predictors are ranked by using Recursive Feature Elimination (RFE). This iterative procedure can be summarized in three recursive steps \citep{guetal02}: the training of the classifier; the computation of the ranking for all features; and the elimination of the feature with smallest ranking.

\section{Results}

\begin{figure}[h]
\centering
 \includegraphics[width=16.5cm]{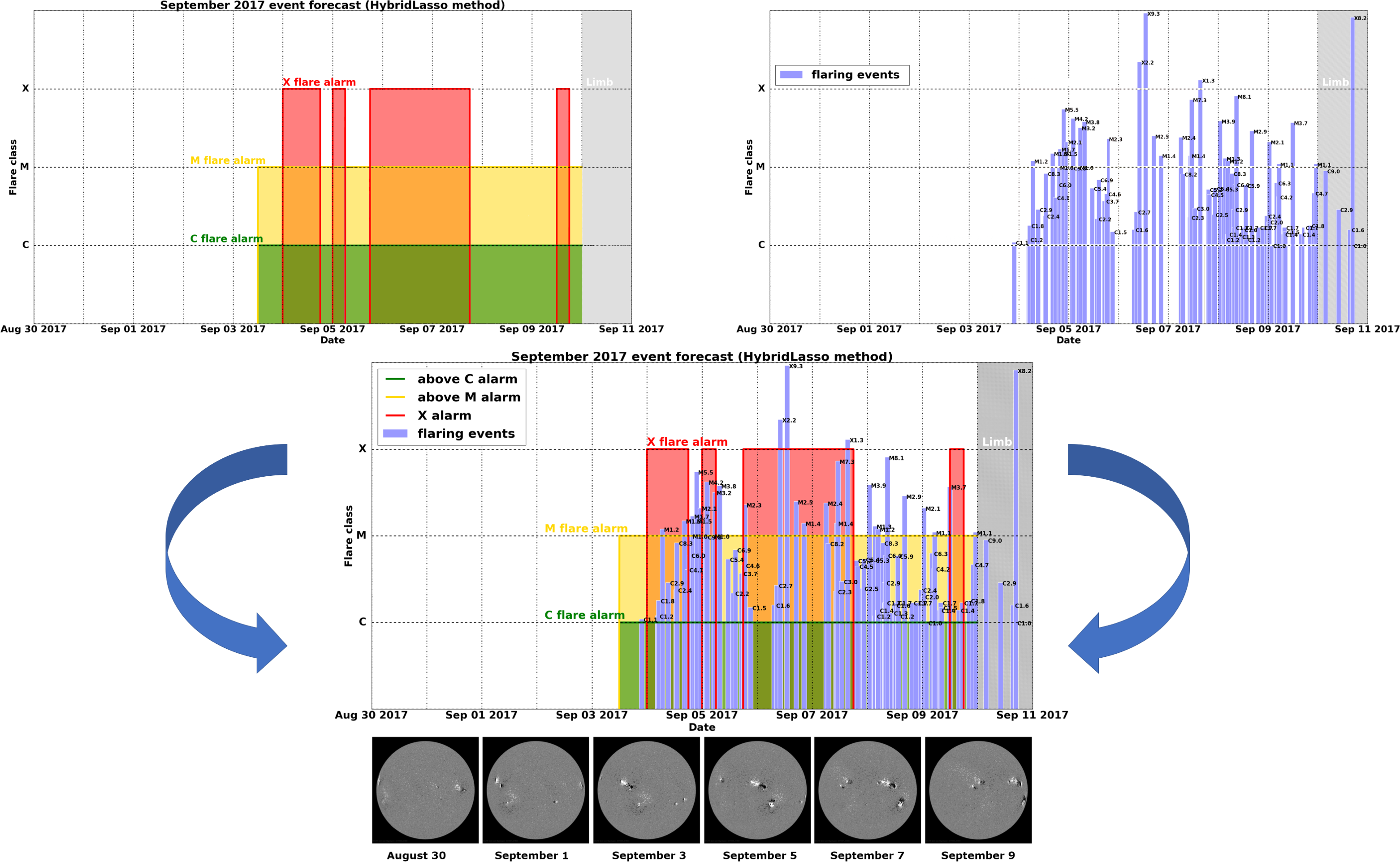}
\caption{Machine learning as a flare warning machine. Top left panel: binary predictions offered along time by the hybrid LASSO algorithm, where green alerts correspond to C1+ flares, yellow alerts to M1+ flares, and red alerts to X1+ flares. Top right panel: actual flaring events recorded by GOES, together with the corresponding GOES flare classes. Bottom middle panel: match between predictions and observations with, at the bottom line, a sample of {\em{HMI}} images used for the analysis.}
\label{figure:fig-2}
 \end{figure}

Once the storm ended, we began assessing the effectiveness of our warning machine and, to this aim, the first, standard thing to do was to compare the alerts sounded by the warning machine with the actual events observed by GOES. Figure \ref{figure:fig-2}, top left panel, puts on a time axis the $24$-hour warnings raised by the algorithm, distinguished with respect to the three flare classes and to the four issuing times: therefore, this panel is the pictorial representation of how the warning machine works. Figure \ref{figure:fig-2} top right panel, contains the actual observations of events recorded by GOES along the same time range, together with the corresponding intensities. If we superimpose the two panels, we obtain, in Figure \ref{figure:fig-2} bottom middle panel, the pictorial representation of the possible matching between predictions and observations. 

What standard practice suggests to do with these data is to compute the corresponding confusion matrices for prediction and observations, and this is what we did in Table {\ref{table:tab-1}}. These numbers show that the confusion matrices for X1+ flares are highly non-diagonal, which can be explained by the notable imbalance of the training set and which systematically implies sub-optimal skill scores values. Yet, these numbers do not fully do justice to the forecasting ability of machine learning. Indeed, given a first sight to what represented in Figure \ref{figure:fig-2}, we can state that:

\begin{enumerate}
\item The warning machine has been completely quiet, until the second part of day September 2, at 12:00:00 UT.
\item At that time point, the machine began sending warnings concerning flares with different intensities.
\item At September 9, 00:00 UT, the algorithm stopped warning (but it must be recalled that on that day AR 12673 exited from the field of view of {\em{HMI}}).
\item GOES began recording flares just at the end of day September 3 and kept on annotating flaring explosions with different flare classes every day, for all days when AR 12673 was visible for {\em{HMI}} and also for the following days. Specifically, two X1+ flares were caught on September 6 and one on September 7.
\end{enumerate}

\begin{figure}
\centering
\begin{tabular}{c}
 \includegraphics[width=11.5cm]{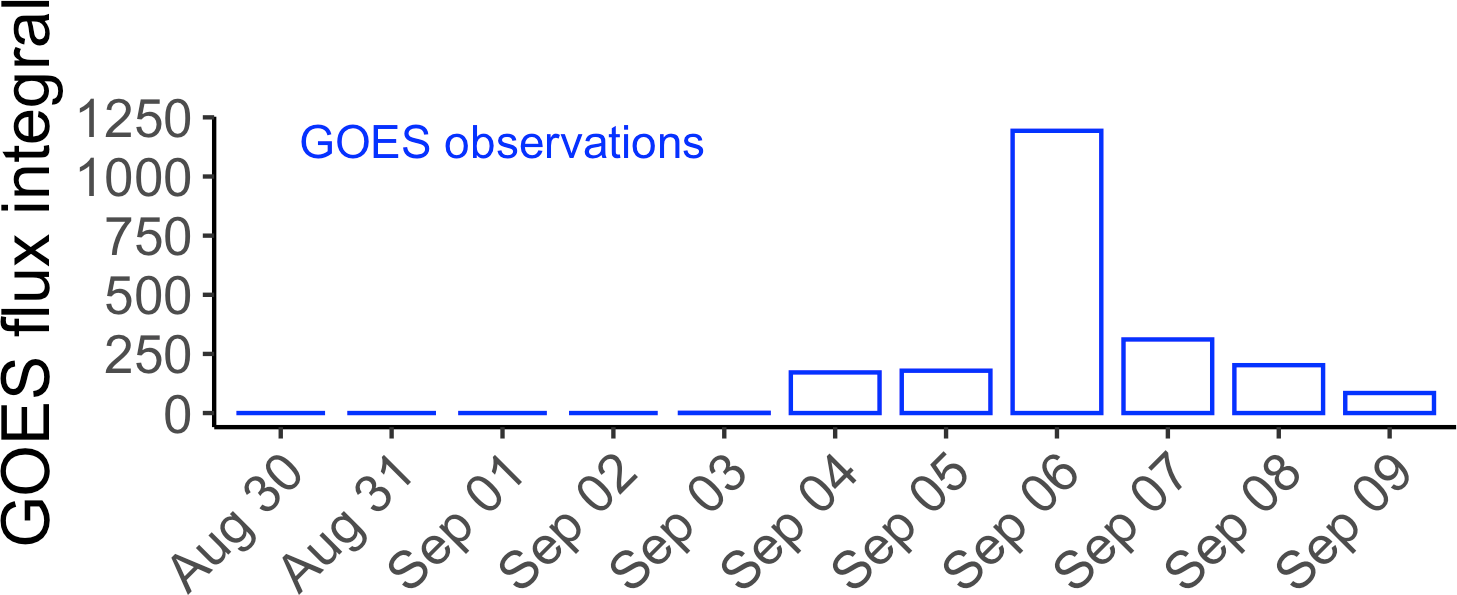} \\
 \includegraphics[width=11.5cm]{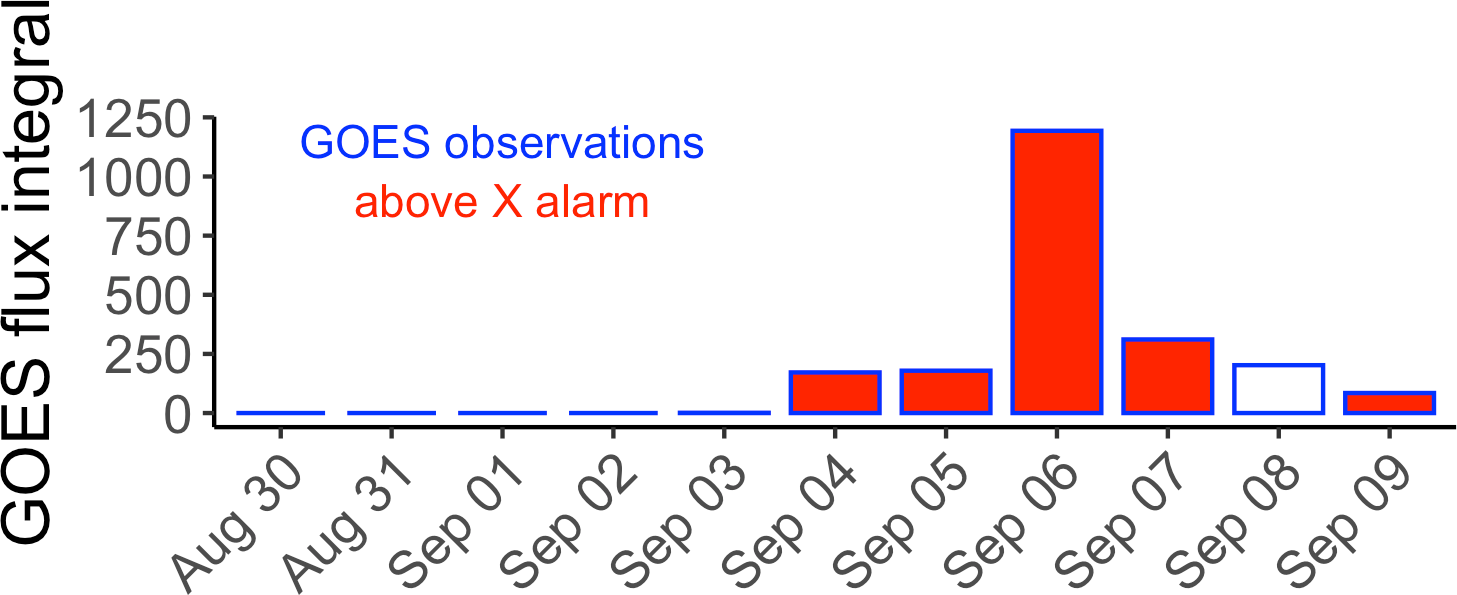}
  \end{tabular}
\caption{Machine learning as a flare warning machine: prediction of the overall amount of energy released at a daily scale. Top panel: the energy budget released day by day according to GOES observations. Bottom panel: comparison between the energy budget released daily according to GOES observations (blue profile) and to lower bound for the energy release as predicted by hybrid LASSO (red filling).}
\label{figure:fig-3}
 \end{figure}

Summing up, this means that the warning machine correctly stayed quiet for all days in which no event actually happened and did not miss a single day in which a flare actually occurred. And, interestingly, at a glance, the {\em{HMI}} images presented at the bottom of the bottom middle panel of Figure \ref{figure:fig-2} and corresponding to different time points, do not seem to present any significant, clearly visibile variation in AR 12673. But, more than this, the warning machine can provide more detailed and more quantitative information if one focuses on the prediction of lower bounds for the energetic budget associated to the emission. In fact, GOES observations clearly witness the huge amount of energy released every day by this AR: considering the overall number of events occurring each day and integrating the daily amount of flux measured by GOES, one obtains (see Figure \ref{figure:fig-3}, top panel) that, each day from September 4 to September 9, the Sun offered an overall activity equivalent to an X1+ event. Coherently with these empirical data, Figure \ref{figure:fig-3}, bottom panel, shows that, with the only exception of September 8, all other days in the time interval from September 4 to September 9, machine learning sounded an alarm for an X1+ flare, thus warning space weather scientists about the occurrence of a notable flare-related energy release within the following $24$ hours.

\begin{figure}
\centering
\begin{tabular}{cc}
 \includegraphics[width=7.5cm]{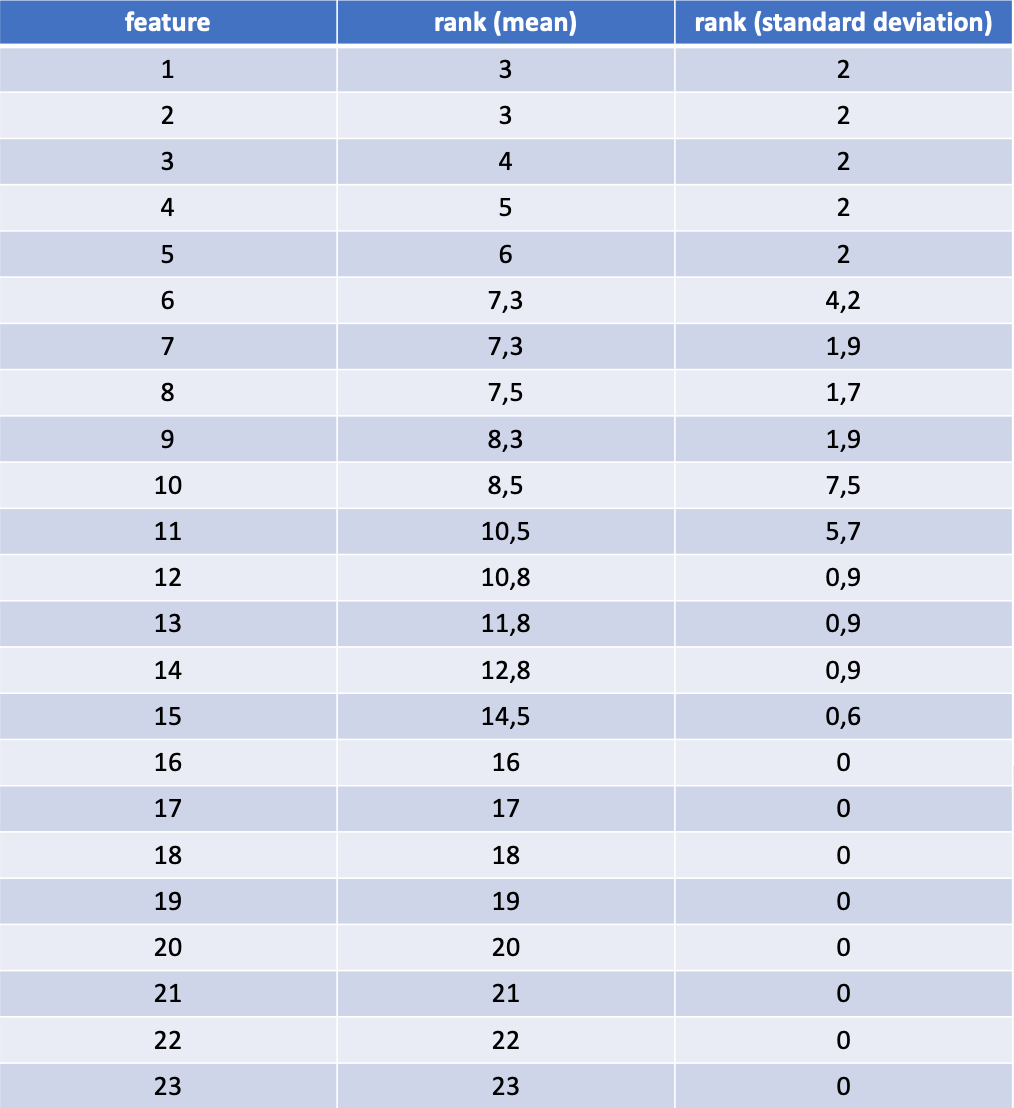} &
 \includegraphics[width=7.5cm]{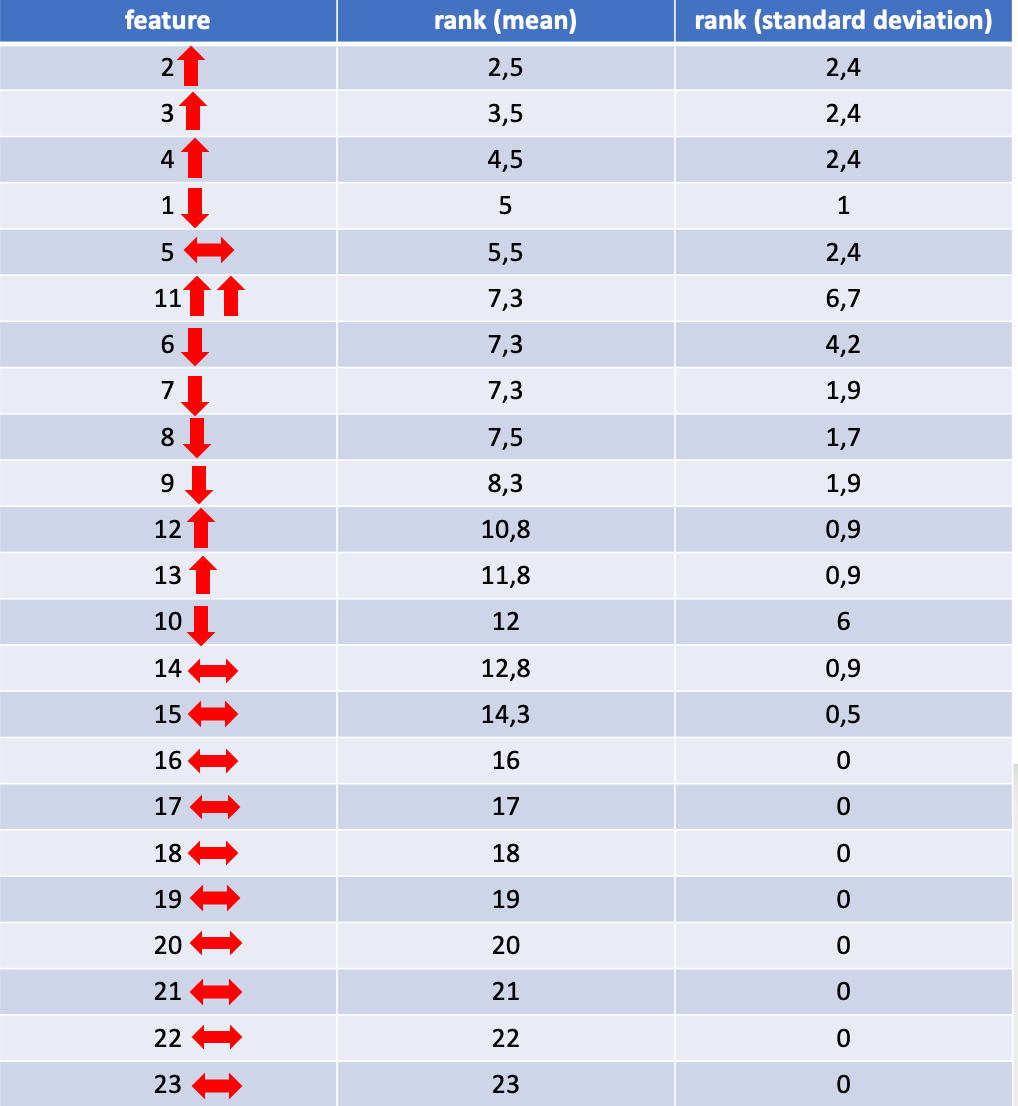}
  \end{tabular}
\caption{Result of the feature ranking process performed by applying RFE on the LASSO regression outcomes. Left panel: the training set is the one used for prediction, i.e. SHARP data in the time range between September 14 2012 and April 30 2016. Right panel: the training set is obtained by enriching the previous one with SHARP data of AR 12673 in the time range between August 30 and September 9 2017. Feature 11 is the only one able to perform five steps forward, from rank 11 to rank 6.}
\label{figure:fig-4}
 \end{figure}

The final part of our study was devoted to investigate the role played by the different features in the prediction process. We applied RFE to the LASSO outcomes twice: first, by considering the same training step utilized for the forecasting obtained in Figure \ref{figure:fig-2}; then, by considering a slightly enriched training set, in which we added also the features extracted from AR 12673 recorded by {\em{HMI}} during the time range from August 30 to September 9 2017. The results of this feature ranking process are illustrated in Figure \ref{figure:fig-4}, which compares the average positions of each feature in the two conditions and the corresponding standard deviations, where average and standard deviation values are computed with respect to the ranks obtained by each feature over the four issuing times considered in the analysis. The two tables in the figure show the notable robustness of the asset of features utilized for training, which is an important confirmation of the reliability of the data analysis process. Interestingly, feature number $11$ is the one characterized by the highest difference in the ranks obtained in the case of the two training sets: specifically, this feature performs five steps forward (from rank 11 to rank 6) when the data concerning the flaring storm are included in the training set. Feature $11$ corresponds to the Ising energy and its increased role in the training process when the storming data are considered seems to reflect the fact that this energetic property contained in the image played a prominent role in the forecasting of an event characterized by the release of a huge amount of stored magnetic energy.

\section{Conclusions}

This study points out that, in the case of flaring events that last for some days, the information extracted by machine learning from {\em{HMI}} data may be accurate enough to setup a warning machine able to sound reliable alerts on a daily scale, in a way similar to what is done by the SWPC-NOAA watches for geomagnetic storms. In September 2017, i.e. toward the minimum of Solar Cycle 24, a huge flaring storm unexpectedly occurred. If this warning machine had been at disposal at that time, then it would have behaved in a very reliable way, staying quiet and sounding warnings coherently with the Sun's activity in the following 24 hours. More than this, the machine would have been able to estimate, at the same daily scale, a lower bound for the total amount of energy released by the flaring events. 

The intrinsic binary nature of the machine, its ability to sound warnings over time and its reliability in predicting the energy budget released during the storm are the three innovative aspects of our system. Just to give a comparison, SolarMonitor relies on probability estimates of the events and their GOES class, and in the case of the September storm these estimates are sometimes not in line with the actual solar activity (by instance, the estimated probability for X flares at September 6 was very low while at that day two X flares occurred).

The machine learning tool utilized in our warning machine is a hybrid LASSO algorithm, i.e. a method that allows the automatic clustering of the regression outcomes in order to realize binary classification and the selection of data features that mostly impact the prediction process. This feature ranking procedure identified the Ising energy as the image property with the highest sensitivity for the training step and this confirms that forecasting methods should consider the energy budget stored in the magnetic field configuration as a crucial parameter of interest. It is therefore probably true that the intrinsic stochasticity of the flaring phenomenon and the notable redundancy of information contained in the observations at disposal prevent satisfactory prediction performances on a point-in-time basis \citep{caetal19}. But it seems also true that the representation of the machine learning outcomes as a set of binary alerts unrolled over time allows the use of the algorithm as a reliable warning machine in the case of flaring storms. And, probably, the result of the feature ranking process suggests that a promising research direction for machine learning nerds is the invention of reliable multi-task flare prediction algorithms that take as input a selection of image properties related to the magnetic energy stored in the magnetograms and provide as output both the binary flare/no-flare classification and some quantitative prediction of the amount of energy actually released during a temporal scale ranging from some hours to a whole day.

\acknowledgements
{This study was supported by the ASI/INAF grant {\em{Artificial Intelligence for the Analysis of Solar Flares Data (AI-FLARES)}}.
}

 \begin{table}
\caption{Confusion matrices concerning the prediction of the September 2017 flaring storm according to the three different flaring classes and the four issuing times considered for the training step.} \label{table:tab-1}
\hskip-3.5cm\begin{tabular}{l | l ||c|c||c|c||c|c||c|c|}
\multicolumn{2}{c}{}&\multicolumn{8}{c}{Flare observed}\\
 \cline{3-10}
 \multicolumn{2}{c}{}&\multicolumn{8}{c}{above C}\\
 \cline{3-10}
\multicolumn{2}{c|}{}&\multicolumn{2}{c||}{00:00:00 UT}&\multicolumn{2}{c||}{06:00:00 UT}&\multicolumn{2}{c||}{12:00:00 UT}&\multicolumn{2}{c|}{18:00:00 UT}\\
\cline{3-10}
\multicolumn{2}{c|}{}&yes&no&yes&no&yes&no&yes&no\\
\cline{2-10}
\multirow{2}{*}{Flare predicted}& yes & 5 & 1 & 6 & 0 & 6 & 1 & 7 & 0\\
\cline{2-10}
&no & 0 & 5 & 0 & 5  & 0 & 4  & 0 & 4\\
\cline{2-10}
\cline{2-10}
\multicolumn{2}{c}{}&\multicolumn{8}{c}{above M}\\
\cline{3-10}
\multicolumn{2}{c|}{}&\multicolumn{2}{c||}{00:00:00 UT}&\multicolumn{2}{c||}{06:00:00 UT}&\multicolumn{2}{c||}{12:00:00 UT}&\multicolumn{2}{c|}{18:00:00 UT}\\
\cline{3-10}
\multicolumn{2}{c|}{}&yes&no&yes&no&yes&no&yes&no\\
\cline{2-10}
\multirow{2}{*}{Flare predicted}& yes & 5 & 1 & 5 & 1 & 5 & 2 & 5 & 2\\
\cline{2-10}
&no &  0 & 5 & 0 & 5 & 0 & 4 & 0 & 4\\
\cline{2-10}
\cline{2-10}
\multicolumn{2}{c}{}&\multicolumn{8}{c}{above X}\\
\cline{3-10}
\multicolumn{2}{c|}{}&\multicolumn{2}{c||}{00:00:00 UT}&\multicolumn{2}{c||}{06:00:00 UT}&\multicolumn{2}{c||}{12:00:00 UT}&\multicolumn{2}{c|}{18:00:00 UT}\\
\cline{3-10}
\multicolumn{2}{c|}{}&yes&no&yes&no&yes&no&yes&no\\
\cline{2-10}
\multirow{2}{*}{Flare predicted}& yes & 0 & 4 & 1& 2 & 1 & 3 & 0 & 2\\
\cline{2-10}
&no & 0 & 7 & 0 & 8  & 0 & 7  & 0 & 9\\
\cline{2-10}
\end{tabular}
\end{table}

\bibliography{astro_sett2017.bib}

\end{document}